# Bifurcations of Emergent Bursting in a Neuronal Network

Yu Wu[*], Wenlian Lu[*], Wei Lin[*], Gareth Leng[†], Jianfeng Feng[*,‡]

[*] Centre for Computational Systems Biology and School of Mathematics,
Fudan University, PR China
[†] School of Biomedicine, Edinburgh University, UK
[‡] Centre for Scientific Computing and Computer Science,
Warwick University, UK
e-mail: jffeng@fudan.edu.cn

June 24, 2011

Abstract

Currently we routinely develop a complex neuronal network to explain observed but often paradoxical phenomena based upon biological recordings. Here we present a general approach to demonstrate how to mathematically tackle such a complex neuronal network so that we can fully understand the underlying mechanism. Using an oxytocin network developed earlier as an example, we show how we can reduce a complex model with many variables to a tractable model with two variables, while retaining all key qualitative features of the model. The approach enables us to uncover how emergent synchronous bursting could arise from a neuronal network which embodies all known biological features. Surprisingly, the discovered mechanisms for bursting are similar to those found in other systems reported in the literature, and illustrate a generic way to exhibit emergent and multi-time scale spikes: at the membrane potential level and the firing rate level.

# 1. Introduction

Oscillatory rhythms in neural system have essential roles in sensory, cognitive, and motor functioning at multiple time scales. As shown in many experimental circumstances[1-3], diverse physiological information can be encoded by the oscillatory activity of neuronal ensembles. The mechanisms of production of rhythmic dynamics vary considerably, from single pacemaker neurons, which can be mathematically described by voltage threshold models, such as the integrate-and-fire model [4], and the more biophysical Hodgkin-Huxley model [5], to large cortical networks, where interactions between neurons are responsible for the rhythmic behaviours ([6] and the references therein).

Single neuron oscillation dynamics are often mathematically interpreted as a dynamic bifurcation, (i.e. an emission of an action potential is regarded a cycle of periodic trajectory). Based on this idea, bifurcation theory has widely been used to investigate neuronal spike dynamics [7]. Conversely, a number of network models have been proposed to realize neuronal oscillation at diverse rhythmic ranges via adapted interactions between inhibition and excitatory neurons [8-10]. Some of these aim to explain the possible roles of several different cortical rhythm ranges ($\delta$ range,1-4 Hz; $\theta$ range,4-8 Hz; $\alpha$ range, 8-13 Hz; $\beta$ range,13-30 Hz; and $\gamma$ range,30-80 Hz) in cognitive functions.

Rhythmic oscillation can be observed and studied at different levels in neural systems, from the single neuron at membrane potential level, to neuronal population at firing rate level. Synchronous spikes in a neuronal population, which is a special case of population oscillating dynamics, may play an essential role in neuronal computation: incognition[11], and attention selection [12,13]in which information is believed to be transmitted in the cortex and/or sub-cortex by synchronous spike chains[14,15]. Synchronization is a population behaviour in a neuronal network, and accordingly has to be studied at the network level. As shown in [16,17], post-synaptic interactions are considered to the cause of the synchronous dynamics. Synchronous bursting emerges periodically in neuronal networks at a time scale of minutes, much longer than the time scale of neuronal spikes, which is in milliseconds. The synchronous behaviour can also be characterized as metastability, that is, a transmission between different patterns [18,19], rather than attractors.

Experimental and computational results above have demonstrated that a neuronal network can exhibit rhythmic oscillations at multi-time scales. An interesting example is reported ina recent paper [20], in which a complex neuronal network was developed to reproduce paradoxical phenomena observed from biological recordings of oxytocin-secreting neurons. Oxytocin is a hormone that is released by neuroendocrine neurons into the blood where it can trigger milk let-down in lactation, and it is also released within the brain, where it has powerful and surprising behavioural effects. Notably, it has effects on social behavior in animals, and in humans it is reported to increase the bonding and trust between individuals. These effects have made oxytocin a key drug target for new therapies aimed at disorders of social behaviour, such as autism.

The neuronal network oxytocin model in [20] was developed to explain the observed activity of oxytocin neurons in response to suckling. When young suckle, they are rewarded intermittently with a let-down of milk that results from reflex secretion of oxytocin; without oxytocin, newly born young will die unless they are fostered.

Oxytocin is made by magnocellular hypothalamic neurons, and is secreted from their nerve endings in the pituitary in response to action potentials (spikes) that are generated in the cell bodies and which are propagated down their axons to the nerve endings. Normally, oxytocin cells discharge asynchronously at 1-3 spikes/s, but during suckling, every 5 min or so, each discharges a brief, intense burst of spikes that release a pulse of oxytocin into the circulation [20]. The near-synchronous bursting is the consequence of vesicles of oxytocin released from the dendrites of oxytocin neurons as a result of spike activity, and this release of oxytocin can activate other oxytocin neurons via its effects on neighbouring dendrites. The model revealed how emergent synchronous bursting at a very low frequency could arise from a neuronal network which implements a number of known features of the physiology of oxytocin cells,. In that model, bursting is an emergent behaviour of a complex system, involving both positive and negative feedbacks, between many sparsely connected cells. The oxytocin cells are regulated by independent afferent inputs, but they are also excited by the dendritic release of oxytocin and inhibited by endocannabinoids, which are also produced by oxytocin neurons as a result of spike activity.

A simple version of the network model is illustrated in Fig. 1. This model network has 48 cells and 12 bundles. Each cell has two dendrites ended up in different bundles, and two cells can interact if they share a common bundle. Each bundle contains the same number of dendrites, which we refer to as a 'homogeneous arrangement of the connections' (Fig. 1A,B). In the model, the dendritic stores of readily-releasable vesicles are continuously incremented by the suckling-related 'priming' input. Their level increases relatively steadily between bursts despite activity-dependent depletion, and synchronous bursts tend to occur when the oxytocin level at the store is relatively high (Fig. 1C,D). In addition to the synchronicity, the bursts possess the characteristic that the inter-burst intervals are almost constant. More interestingly, we observed a number of paradoxical behaviours. For example, increased spike activity between the burst enhances depletion of the stores and so can delay or even suppress bursting (Fig. 1E).Conversely, an increase in inhibitory inputs can promote the reflex in a system which fails to express bursting because of insufficient priming. For example, injections of the inhibitory neurotransmitter GABA into the supraoptic nucleus of a suckled, lactating rat can trigger milk-ejection bursts (Fig. 1F).

This neuronal network illustrates a hierarchical rhythmic oscillation dynamics: each neuron emits action potentials periodically that can be regarded as oscillating dynamics at neuron level (the msec time scale for the inter-spike-interval); the network population synchronizes and exhibits bursting dynamics periodically that can be regarded as oscillating dynamics at network level (the minute time scale for the inter-burst-interval), comparing Fig. 1C with Fig. 1D. In general, a network system can have diverse oscillation dynamics at different levels, owing to the interactions between individual units. Each node oscillates and exhibits a faster rhythmic dynamics. The network synchronization also oscillates and shows a slower rhythmic dynamics.

Different approaches have been proposed to deal with hierarchical rhythmic dynamics, as exemplified by neuronal bursting. The theory of slow-fast dynamical systems was introduced to explain how a neuron model can demonstrate co-existence of tonic spiking and bursting [21,22]. Abundant bifurcation behaviours in oscillations including spiking and bursting were detected in various neuron models [23-28], and are thought to be biophysically plausible mechanisms. Besides, reduction of complex neuronal networks to models with a few variables was performed and mean field models were constructed to describe the average activity of the neuron systems [22,29-31].

Here, we aim to explain why and how emergent bursting occurs in the oxytocin network, and to reveal the underlying mechanisms of the puzzle: how increasing excitatory inputs can sometime stop the burst and increasing inhibitory inputs can promote the burst. Despite the large number of published papers in this area, we find that a novel approach is still required. First, most theories only deal with deterministic dynamics. However, in our model, each neuron receives stochastic (Poisson) inputs, so an approximation to simplify each single neuron model is needed. We approximate it by 'the usual approximation'. Second, we approximate the system by a two-dimensional dynamical system: a slow-fast dynamical system, where the variables used are threshold and oxytocin store level. This simplification is achieved after intensively testing different model variables. The original model included many variables that were needed to match the physiological data quantitatively, including a hyperpolarizing after potential (HAP) and a slow afterhyperpolarising potential, different delays in the systems, and variables to model endocannabinoids actions. This complexity makes the original model hard to deal with mathematically. After omitting non-essential variables, while keeping the key features of the model, we conclude that a model incorporating just the dynamics of the threshold and oxytocin store level can be used to mimic the original model. Using the two dimensional model, we apply bifurcation theory to deal with the hierarchical rhythmic dynamics. Precisely, we find there exists a critical value of the input rate beyond which bursting can emerge. This phenomenon can be perfectly described by a *saddle-node bifurcation of limit cycles*. As the excitatory inputs increase in frequency, synchronized bursts arise in such a manner that the intervals between each burst are identical. More interestingly, and counter intuitively, the bursts disappear when the excitatory input frequency passes a larger critical value corresponding to another saddle-node bifurcation of limit cycles. We also detect occurrences of the subcritical Hopf bifurcation as the input frequency varies between the above two critical values. The saddle-node bifurcation plays a more significant role corresponding to the generation and ending of the bursting activity in the network.

## 2. Models/Methods

*2.1 Oxytocin neuronal network*

In [20], a neuronal network model, based on leaky integrate-fire neuron and with adaptive thresholds, dependent on the storage level of oxytocin, was proposed and proved to exhibit emergent bursting dynamics. In this model, neurons are subject to random synaptic inputs from other neurons(excitatory and inhibitory post synaptic potentials, EPSPs and IPSPs). Changes in excitability of the neurons were modeled as changes in the membrane potential threshold for emission of action potentials, and depend on the spike trains and on dendritic oxytocin release, which is non-linearly related to spike activity and proportional to the size of the readily-releasable store of oxytocin in the dendrites. As a closed loop, the stores of oxytocin that are available for release decrease when oxytocin is released from the dendrites but are increased as a result of the suckling stimulus. In the current paper we consider a simplified version of the bursting neuronal network model, which still preserves the synchronous bursting behaviour. We refer the model in [20] as the original neuronal network (ONN) and our simplified model as the simplified neuronal network (SNN).

The core step of our simplification is the topology of the network. We consider a neuronal network with $N$ neurons and $n_b$ bundles, where each neuron has two dendrites in different bundles. We assume that the network is homogeneously arranged, i.e. each of the $n_b$ bundles contains the same number of dendrites. In [20], we modeled the individual oxytocin neurons using the leaky integrate-and-fire model, modified to incorporate activity-dependent changes in excitability. The membrane potential $v_i$ of cell $i$ obeys

$$\frac{dv_i}{dt} = \frac{v_{rest} - v_i}{\tau} + \sum_{j=1}^{2}\left[a_E(v_E - v_i)\frac{dN_{E,i}^j}{dt} - a_I(v_i - v_I)\frac{dN_{I,i}^j}{dt}\right], \qquad (1)$$

where $\tau$ is the membrane time constant, $v_{rest}$ is the resting potential, $N_{E,i}^j, N_{I,i}^j$ are independent Poisson processes with the varied excitatory input rate $\lambda_{E,i}^j$ and the fixed inhibitory input rate $\lambda_{I,i}^j$, $a_E(v_E - v_{rest}), a_I(v_{rest} - v_I)$ are the magnitude of single EPSPs and IPSPs at $v_{rest}$, and $v_E, v_I$ are the excitatory and inhibitory reversal potential. A spike is produced in cell $i$ at time $t = t_i^s, s = 1,2,...$, if $v_i(t_i^s) = T_i(t_i^s)$, where $T_i(t)$ is the spike threshold at time $t$. After a spike, $v_i$ is reset to $v_{rest}$. Activity-dependent changes in excitability and the effects of oxytocin are modeled by effects on spike threshold. Different from the model for the dynamical threshold in [20], we eliminate the effects of HAP and AHP in the spike threshold, that is,

$$T_i(t) = T_0 - T_{OT,i}(t),$$

where $T_0$ is a constant. The increase in excitability due to oxytocin is modeled by $T_{OT}$,

$$\frac{dT_{OT,i}}{dt} = -\frac{T_{OT,i}}{\tau_{OT}} + k_{OT} \sum_{k=1}^{n_b} \sum_{j=1}^{N} \sum_{l,m=1}^{2} c_{il}^k c_{jm}^k \rho_j^m(t), \tag{2}$$

where $\tau_{OT}, k_{OT}$ are constants, $\rho_j^m(t)$ is the instantaneous release rate from dendrite $m$ of cell $j$, and the sums pick up all the cells whose dendrites share the same bundle as cell $i$. The network topology is represented by matrices $C^k = \{c_{ij}^k\}, k = 1,...,n_b; c_{ij}^k = 1$ if dendrite $j$ of cell $i$ is in bundle $k$, and zero otherwise.

The readily-releasable store of oxytocin in dendrite $j$ of cell $i$ is represented by $r_i^j$, where

$$\frac{dr_i^j}{dt} = -\frac{r_i^j}{\tau_r} + k_p - \rho_i^j(t), \tag{3}$$

where $\tau_r$ is a time constant, $k_p$ is the rate of priming due to the suckling input, and $\rho_i^j$ is the instantaneous release rate from dendrites $j$. In [20], the release of oxytocin is proportional to the readily-releasable stores:

$$\rho_i^j(t) = k_r r_i^j(t) \sum \delta(t - t_i^s - \Delta), \tag{4}$$

where $k_r$ is the maximum fraction of the stores that can be released by a spike, $\Delta$ is a fixed delay before release, and the summation extends over the set $\{t_i^s < t, t_i^s - t_i^{s-1} < \tau_{rel}\}$, with $\tau_{rel}$ a constant. This ensures that only spikes occurring at intervals of less than $\tau_{rel}$ induce any release from dendrites. Here, we neglect the delay term $\Delta$ in (4) and the doublet effects by letting $\tau_{rel} = +\infty$, which means that spikes occurring at intervals of any length can induce release.

The model in [20] also took the inhibitory effects of endocannabinoids into consideration, but here we neglect it for simplification.

The parameter values for simulations are as in Table 1.

The ONN in [20] displays the transition between spiking and bursting (Fig. 2). The spiking rate is recorded on a network of 48 neurons and 12 bundles in Fig. 2A, and the voltage trace and store level of oxytocin are shown in Fig. 2C and E. Note that the bursting events are essentially attributed to the drop of the spiking threshold (red line) and store level. Our simplification of the ONN does not destroy such basic behaviours of the network in the sense that the SNN displays similar network activity in Fig. 2B, 2D and 2F as the ONN in Fig. 2A, 2C and 2E. As expected, the SNN fires faster than the ONN even though the input rate $\lambda_{E,i}^j$ in the SNN (50 Hz) is smaller than in the ONN (80 Hz), because we have discarded all bursting terminating mechanisms related to the negative feedbacks of HAP and AHP in the spike threshold, the doublet effects in the impulsive release of oxytocin and the inhibition of endocannabinoids.

Next we regard $t_i^s$ as a series of random variables, and then the usual approximation of the release rate takes the following form:

$$\rho_i^j(t)dt = k_r r_i^j(t)[\mu_{output,i}(t)dt + \sigma_{output,i}(t)dB_i(t)]. \qquad (5)$$

where $\mu_{output,i}$ is the spiking rate and $\sigma_{output,i}$ is the variance of the correlated Brownian motions $B_i(t), i = 1,...,N$.

Because of the assumption that the network is homogeneously arranged and the observation that the neuronal population is activated synchronously, it is a useful approximation by employing the mean field method. Explicitly, let $T_{OT}, r$, and $\mu_{output}$ denote the corresponding dynamical variables averaged over the entire population, and suppose that the number of entities in the summation in (2) is $n$ ($n = 4N/n_b$). As a first approximation, we can ignore the random effect and then omit all the subscripts in $T_{OT,i}, r_i^j, \mu_{output,i}$ in (2), (3) and (5) respectively. A two dimensional determinant dynamical system that describes the behaviour of the averaged neuronal activity is as follows:

$$\begin{cases} \dfrac{dr}{dt} = -\left(\dfrac{1}{\tau_r} + k_r \mu_{output}(t)\right) r + k_p, \\ \dfrac{dT_{OT}}{dt} = -\dfrac{T_{OT}}{\tau_{OT}} + k_{OT} k_r n \mu_{output}(t) r. \end{cases} \qquad (6)$$

We make a further simplification by removing the limitation on the maximal value of the reduction of the spike threshold which is set to be 25 mV in [20].

## 2.2 Firing rate map approximation

In the system (6), $\mu_{output}(t)$ is an unknown term varying with time, which makes (6) a non-autonomous system. To overcome the difficulty, here we present a method to evaluate the mean firing rate $\mu_{output}(t)$ of the network activity so that the system (6) becomes a mathematically tractable autonomous system. Intuitively, the firing rate $\mu_{output}(t)$ varies in response to the fluctuation of the spike threshold $T(t)$ and the frequency of the afferent input $\lambda_E$. If we write the firing rate $\mu_{output}(t)$ as a function of the time-varying threshold $T$ and the input rate $\lambda_E$:

$$\mu_{output} = \mu_{output}(T, \lambda_E), \qquad (7)$$

a firing rate map, and substitute (7) in (6), we obtain the following two-dimensional system with two parameters $\lambda_E, n$:

$$\begin{cases} \dfrac{dr}{dt} = -\left(\dfrac{1}{\tau_r} + k_r \mu_{output}(T_0 - T_{OT}, \lambda_E)\right) r + k_p, \\ \dfrac{dT_{OT}}{dt} = -\dfrac{T_{OT}}{\tau_{OT}} + k_{OT} k_r n \mu_{output}(T_0 - T_{OT}, \lambda_E) r. \end{cases} \qquad (8)$$

To find the analytical expression of the firing rate map, we adopt a numerical approach by simulating the leaky integrate-fire model. Simulations of equation (1) for a single cell are conducted by fixing $T$ on each trial. Fig.3A shows the relationship between $\mu_{output}$ and $T$ corresponding to varied excitatory inputs.

Due to the shape of the firing rate map in Fig.3A, we use a sigmoid-like function to fit it:

$$\mu_{output}(T,\lambda_E) = \frac{1000}{1+\exp\left\{\frac{T-\alpha(\lambda_E)}{\beta(\lambda_E)}\right\}} + \gamma(\lambda_E).$$

Here $\alpha(\lambda_E)$ is the center of the curve, and $\beta(\lambda_E)$ is a tunable factor that controls the sharpness, $\gamma(\lambda_E)$ is the term to describe the spike activity when the spike threshold is at the initial level $T_0 = -50$ Hz. By numerical experiments, we find $\alpha(\lambda_E) = -66 + 0.02\lambda_E$, $\beta(\lambda_E) = \sqrt{0.02(\lambda_E + 20)}$ and $\gamma(\lambda_E) = 35\left(\frac{\lambda_E}{200}\right)^{\frac{5}{2}}$. Fig. 3B shows the plot of the constructed function $\mu_{output}(T,\lambda_E)$.

To summarize all procedures above, we here include a flow chart as in Fig. 4 to illustrate our endeavors. In the first step, we simplify a network model with a single neuron of 10 variables by discarding the negative feedbacks in the spike threshold and the doublet effects on the impulsive release of oxytocin, and obtain a simplified network model with 4 variables for each neuron. After evaluating the firing rate map, we derive the reduced deterministic autonomous system (8) (also called the mean field model) in the second step, which enables us to perform the bifurcation analysis. The similar approach could be employed as a general guidance to deal with other complex and stochastic neuronal networks.

## 3. Results

### 3.1 Bifurcation analysis

It can be observed that the value of the dynamical spike threshold $T$ is closely related to the appearance of the bursting behaviours. In particular, a lower value of $T$ can trigger a burst. Therefore, a systematic exploration of the dynamical properties of the system (8) enables us to understand the mechanism of the entire network activities.

Allowing the parameter $\lambda_E$ (resp. $n$) to vary while keeping the other parameter $n$ (resp. $\lambda_E$) fixed, the system (8) displays two types of bifurcations: the saddle-node bifurcation of limit cycles and the subcritical Hopf bifurcation. To exemplify this conclusion of the bifurcations in Eq. (8), we first investigate the system dynamical behaviours by fixing $n = n_0 = 22$ and varying the input $\lambda_E$.

When the value of $\lambda_E$ is small, the unique fixed point equilibrium in the $r$-$T_{OT}$ plane is asymptotically stable. Thus, from the asymptotical convergence of the trajectory if $[r(t), T_{OT}(t)]$, as shown in Fig. 5A, we conclude that there is no bursting activity.

When $\lambda_E$ increasingly passes a critical value $\lambda_E^c \approx 60.1386343160437030$, the saddle-node bifurcation of limit cycles occurs. To demonstrate the existence of this bifurcation and verify the stability of the bifurcated limit cycle, we construct the Poincaré map of $(8)|_{\lambda_E}$. Denote the equilibrium point by $x_0(\lambda_E) = [r_0(\lambda_E), T_{OT,0}(\lambda_E)]$ and set

$$L_{\lambda_E} = \{[r_0(\lambda_E), T_{OT}] \mid T_{OT} \leq T_{OT,0}(\lambda_E), T_{OT} \in R\},$$

so that $L_{\lambda_E}$ is a half line transversal to the vector field in the neighborhood of the equilibrium $x_0(\lambda_E)$. Here we introduce a new coordinate system along $L_{\lambda_E}$, where $x_0(\lambda_E)$ is seen as an origin and $n_{\lambda_E}$ is a unit vector parallel to $L_{\lambda_E}$. Hence, $\alpha$ becomes the coordinate of a point $x$ on $L_{\lambda_E}$ if $x = \alpha n_{\lambda_E} + x_0(\lambda_E)$ for some $\alpha \geq 0$. Now, suppose that $\phi_{\lambda_E}(t,x)$ represents the solution of $(8)|_{\lambda_E}$ with the initial point $x$. Mathematically, it can be validated that there exists a number $\bar{t} = \bar{t}(x) > 0$ such that $\phi_{\lambda_E}(\bar{t}(x),x) \in L_{\lambda_E}$ and $\phi_{\lambda_E}(t,x) \notin L_{\lambda_E}$ for $t \in (0, \bar{t}(x))$. In other words, $\phi_{\lambda_E}(\bar{t}(x),x)$ is the point at which the trajectory intersects with $L_{\lambda_E}$ for the first time after it departures from the initial point $x$. Thus, the coordinate $\bar{\alpha}$ of $\phi_{\lambda_E}(\bar{t}(x),x)$ can be uniquely determined through $\phi_{\lambda_E}(\bar{t}(x),x) = \bar{\alpha} n_{\lambda_E} + x_0(\lambda_E)$, and consequently the Poincaré map, denoted by $P: R^+ \to R^+$, is established by $P(\alpha) = \bar{\alpha}$ for $\alpha > 0$. Fig. 6B shows the curves of the constructed Poincaré map $P$ for different values of $\lambda_E$, where, clearly, each intersection between the curves and the black line $P_1(\alpha) = \alpha$ is a fixed point of $P$. When $\lambda_E$ is smaller, $P$ has no fixed point for $\alpha \in R^+$. When $\lambda_E = \lambda_E^c$, it has a unique fixed point. Since the quantity $1 - |P'(\alpha)|$ at the two sides of the fixed point has different signs, this fixed point is attracting on the right side and repelling on the left. When $\lambda_E$ becomes slightly larger than $\lambda_E^c$, two fixed points branch off: one is stable and the other is unstable. These stabilities can be derived from the sign of the above quantity at different fixed points. For example, when $\lambda_E = 61\,\text{Hz}$, the quantities at the two fixed points are $0.94$ and $-4.06$, respectively. Furthermore, since the fixed points of $P$ correspond to limit cycles, the system $(8)|_{\lambda_E = \lambda_E^c}$ has a semi-stable limit cycle and the system $(8)|_{\lambda_E \geq \lambda_E^c}$ has two bifurcated limit cycles: the one with a larger amplitude is stable and the other in the interior is unstable. In the simulation, the two bifurcated limit cycles can be numerically observed (Fig. 5B).

As shown in Fig. 5C-D, the interior limit cycle gradually shrinks to the equilibrium as $\lambda_E$ increasingly departs from $\lambda_E^c$ to $\lambda_E^{h_1} \approx 64.9\,\text{Hz}$. When $\lambda_E$ passes through $\lambda_E^{h_1}$, a so-called subcritical Hopf bifurcation occurs. The stable limit cycle is preserved, but the shrinking interior limit cycle coincides with the equilibrium, and this makes the equilibrium become unstable (Fig. 5E). Here, it is worth mentioning that the stabilities of the equilibrium and the limit cycle attributed to the Hopf bifurcation can be validated by calculating the first Lyapunov coefficient (FLC). The FLC for the bifurcation point $\lambda_E^{h_1}$ is $0.4721 > 0$, which validates the existence of the subcritical Hopf bifurcation.

Interestingly, aside from the above two bifurcations, the other two bifurcations appear almost symmetrically and consecutively. In fact, when $\lambda_E$ passes through $\lambda_E^{h_2} \approx 90.9$ Hz, the other subcritical Hopf bifurcation of the system (8) emerges with a positive FLC, (0.6262), which results in the change of the stability of the originally-unstable equilibrium and brings an unstable limit cycle (Fig. 5F-G). Moreover, the amplitude of the bifurcated unstable limit cycle grows until $\lambda_E$ increasingly approaches $\lambda_E^n \approx 99.6$ Hz, where the other saddle-node bifurcation occurs. This bifurcation leads to the coalescence and annihilation of the two limit cycles (Fig. 5H-I). The above-expatiated bifurcation procedure of the system (8) is illustrated in Fig. 7A.

As mentioned above, a burst is triggered if the spike threshold $T$ is sufficiently low. Also, because of the bifurcated stable limit cycle, there exists a stable periodic obit fluctuating between the two critical excitation levels. This indicates that burst events can occur continuously and that the inter-burst interval is equal to the period of the stable periodic orbit, which illustrates the invariance of the inter-burst interval. Fig. 7B dynamically shows the spiking threshold and store level of the system (8) as the parameters are taken as $\lambda = 90.9$ Hz and $n = n_0$. Since $\tau_r$ is always set as a value much larger than $\tau_{OT}$, the sharp peaks in Fig. 7B (left) and the sharp valleys in Fig. 7B (right) reflect the characteristics of the slow-fast dynamical system.

For $\lambda_E > 99.6$, the system (8) has no limit cycle but only one stable fixed point. In such a case, the dynamical behaviour of the system is analogous to that of the system with a small value of $\lambda_E$. Therefore, bursts disappear as excitation is beyond the critical level. From the aspect of the ONN, oxytocin released events may emerge so frequently that the stores are not replenished fast enough to reach the critical level required to trigger a burst. Under such conditions, bursts are rarer and less predictable, until eventually over-excitation disrupts the reflex secretion of oxytocin [20].

In Fig. 7C, the phase trajectories of the storage level and $T_{OT}$ are plotted to show the bifurcation transition regulated by the input rate $\lambda_E$. Here, we fix the parameter $n$ at 22. As shown in the inner plot of Fig. 7C, we start with stable attractor (the green star) with $\lambda_E = 57$ Hz. By increasing the input rate to 62 Hz that is located in the bifurcation region as shown in Fig. 7A, the phase trajectory goes to a limit cycle (the purple curve). The bifurcation of bursting is generated. Keeping increasing the excitation input to the network so that the rate enters the high-rate stable attractor region as shown in Fig. 7A, the system becomes stable again (the red curve and star). That is, the bursting activity is destroyed by overwhelm excitation inputs. By decreasing the excitatory input rate to the bifurcation region, which is equivalent to increasing the inhibitory input rate, the system goes to a limit cycle (the blue curve) so that the bursting is induced. This transition coincide with the phenomena shown in Fig. 1F that injections of the inhibitory substance can trigger bursts.

**3.2 Comparing the mean field model with the SNN**
Based on the bifurcation analysis, we return to the network bursting dynamics and compare the SNN and the mean field model (Fig. 6B). In the SNN, a burst is recorded if the firing rate is larger than 30Hz. For a network of 48 neurons and 12 dendrites, bursting emerges when the excitatory input frequency is between 15Hz and 70Hz. In the mean field model, we can replicate the network bursting dynamics from the traces of voltage and store level. For given $n$ and $\lambda_E$, we say that there is a network bursting if a stable limit cycle exists in the reduced system, and the inter-burst interval is the period of the limit cycle. Therefore, the burst frequency in the mean filed model shown in Fig. 7D is the reciprocal of the period of the limit cycle. For the SNN with a typical population size of 48 neurons and 12 dendrites, the value of $n$ in the corresponding mean field model should be $n = 4N/n_b = 16$. To compare with the SNN, here we pick $n = 22$ in the mean field model for the reason stated in the next section. Fig. 7D shows that the replicated 'network' possesses similar bursting dynamics to the SNN.

# 4. Discussions

In the current paper, we presented a general approach to tackle a complex neuronal network dynamics which exhibits spikes at multi-time scales. Our approach allows us to simplify a network model with a single neuron of 10 variables to a simple two dimensional model: the mean field model. The approach is general in the sense that it can be easily applied to dealing with other similar neuronal networks. As mentioned before, the oxytocin model has attracted considerable interests in the literature and other groups have tried to investigate analytically as well [22]. Their approach is interesting, but is quite far off from revealing the actual mechanisms of the model. Another close related model is presented in [32] and its dynamical behavior should be very similar to ours, as pointed out in our early paper [20], although we have not seen published work on it [33].

The complex neuronal network introduced in the earlier paper [20] possesses a hierarchical rhythmic structure. Under the homogenous topological assumption of the network, the neurons display spiking activities induced by afferent inputs at the neuronal level, while the global network demonstrates synchronous oscillation at the network level. The ONN showed paradoxical network behaviours that the bursting events occur continually when the excitatory input rate is at a certain range, but disappear when the excitatory input rate is sufficiently large.

Based on the ONN, we developed a simplified version of the neuronal network (SNN) which preserves these basic bursting behaviours. Then, we utilized the mean field approach and reduced the SNN to a mean field model, and the bursting neuronal activity corresponds to a limit cycle in the mean filed model. The critical step in the approach above is the firing rate map approximation. We obtained the map via numerically simulating the leaky integrate-fire model with fixed threshold in each trial. A sigmoid-like function is then constructed to approximate the firing rate map.

The mean field model is a two-dimensional dynamical system with two parameters, $\lambda_E$ and $n$, where $\lambda_E$ is the excitatory input frequency and $n$ denotes the connection strength. For a fixed $n \geq 22$, we have found that the dynamical system (8) displays two types of bifurcations as $\lambda_E$ varies: a saddle-node bifurcation of limit cycle and a subcritical Hopf bifurcation. The former bifurcation accounts for the generation and ending of the bursting events and the identical inter-burst intervals, which is more significant in the network behaviours. However, for $n < 22$, there is only one fixed point remaining in the dynamical system, indicating that burst no longer exists (see below).

In the preceding investigations, bifurcations are studied with the variable $n$ fixed at 22. Actually, $n$ is determined by the scale and the connection of the network. Numerical investigations of the system (8) show that, if $n$ is small, there is no bifurcation for any possible values of $\lambda_E$. Indeed, for $n < n_0 = 22$, the system (8) has no limit cycles, but a unique stable equilibrium, i.e. no bursting activity of networks appears in such a case. For a given $n \geq n_0$ and with the variation of $\lambda_E$, a pair of conjugate eigenvalues of the linearized system transversally cross the imaginary axis twice, so that the limit cycle generated by the Hopf bifurcation emerges. This further makes it possible to generate stable limit cycles coexistent with unstable limit cycles. This also explains why we picked $n = 22$ for the mean field model in the comparison with the SNN with $n = 16$ (see Fig. 7D).

Actually, the bifurcations with respect to the input rate and network size can be summarized as the emergence of a codimension two bifurcation, namely the Bautin bifurcation, by regarding the mean field model as a member in the two-parameter family of autonomous ordinary differential equations. This bifurcation is beyond the scope of this paper.

We should point out that the phenomena based on the bifurcations of network size described above for the mean field model are not consistent with the ONN or SNN. As for the SNN as well as the ONN, even with a small network population or weak connection (i.e., each neuron is connected with few other neurons), bursting events still exist. Actually the mean field model might be more reasonable and closer to the underlying mechanisms of the real neuronal system in the sense that bursts could hardly be triggered for a single or few neurons. The discrepancy between the mean field model and the ONN tells us the shortcomings of the ONN model, despite of the successfully fitting of the model with experimental data.

The main purpose of a mathematical model is to reveal the true mechanism of a complex biological system, while at the same time retaining its main features. It is certainly unsatisfactory if we can only replace one complex (biological) system with another equally complex (mathematical) system. The ultimate aim of a mathematical model is to capture the essence of the system so that we can understand, interfere and control the system. By reducing the original complex system to a two dimensional system, we successfully grasp the property of the model and biological network. The approach adopted here can certainly be generalized to analyze other systems.

# Figure captions:

Figure 1. Oxytocin network and its behaviour at multi-time scales. (A) Schematic diagram illustrating the topology of the model network; for each cell, two yellow squares indicate which bundles are occupied by the cell dendrites. (B) A few clusters of cells are found in the network where neurons (circles) interact via both dendrites (lines). Such clusters may occasionally be connected through a common bundle. (C) Ratemeter records of 3 representative cells showing bursts in response to simulated suckling. A clear spike at the firing rate level is observed. (D) Raster plots of the activity of all 48 cells in the network through the first simulated milk-ejection burst. Note the approximately synchronous activation of all model cells during a burst. (E) Adding excitatory input to the network will paradoxically destroy the bursting activity. (F) Increasing inhibitory input can sometime induce the bursting.

Figure 2. Transition between spiking and bursting in the ONN with $\lambda_E = 80\,\text{Hz}$ (left column) and in the SNN with $\lambda_E = 50\,\text{Hz}$ (right column). Both networks are composed of 48 neurons and 12 bundles. (A,B) Ratemeter records of 5 representative cells with time span of 600s. (C,D) Voltage trace(blue) and spiking threshold(red) of cell 1. (E,F) Records of the store level of oxytocin of cell 1. Note that the bursting events are essentially attributed to the drop of the spiking threshold and store level.

Figure 3. (A) The relationship between $\mu_{output}$ and $T$ corresponding to varied $\lambda_E$. (B) Plot of the constructed function $\mu_{output}(T, \lambda_E)$.

Figure 4. A flow chart for the illustration of our endeavors. In the first step, we simplify a network model with a single neuron of 10 variables by discarding the negative feedbacks in the spike threshold and the doublet effects on the impulsive release of oxytocin, and obtain a simplified network model with 4 variables for each neuron. After evaluating the firing rate map, we derive the mean field model in the second step, which enables us to perform the bifurcation analysis.

Figure 5. Phase portraits of the system (8) with $n = n_0$ and different $\lambda_E$: (A) $\lambda_E = 20\,\text{Hz}$; (B) $\lambda_E = 60.13865\,\text{Hz}$; (C) $\lambda_E = 60.2\,\text{Hz}$; (D) $\lambda_E = 63\,\text{Hz}$; (E) $\lambda_E = 64.9\,\text{Hz}$; (F) $\lambda_E = 90.9\,\text{Hz}$; (G) $\lambda_E = 92.5\,\text{Hz}$; (H) $\lambda_E = 99.6\,\text{Hz}$; (I) $\lambda_E = 99.7\,\text{Hz}$. Here, the $r$-nullcline and $T_{OT}$-nullcline are colored in blue and green, respectively. The red circles represent the unstable limit cycles, and the black curves stand for the orbits with the initial point $(0,0)$.

Figure 6. (A) Establishment of a coordinate system on the half line $L_{\lambda_E}$ with the origin $x_0(\lambda_E)$. Here, $x_0(\lambda_E)$ is the equilibrium point and $L_{\lambda_E}$ is transversal to the vector field in the neighborhood of $x_0(\lambda_E)$. Also note that both $x_0(\lambda_E)$ and $L_{\lambda_E}$ depend continuously on $\lambda_E$; (B) Curves of the Poincaré map $P$. Each intersection

between the curves and the black line $P_1(\alpha) = \alpha$ corresponds to a fixed point of $P$ as well as to a limit cycle of the system (8). Particularly, for $\lambda_E = 59\,\text{Hz}$, the curve has no intersection with the black line, so that there is no limit cycle. With the increase of $\lambda_E$, the curve moves upward. It first intersects with the black line at $\lambda_E = \lambda_E^c$, where a single semi-stable limit cycle emerges. As $\lambda_E$ increases to $61\,\text{Hz}$, two bifurcated limit cycles appears. Here, one cycle is stable characterized by the quantity $1-|P'(\alpha)|= 0.94 > 0$ at one fixed point and the other cycle is unstable with the quantity $-4.06 < 0$ at the other fixed point.

Figure 7. Illustration of the bifurcation behaviours in the mean field model. (A) Bifurcation diagram with $n = n_0$ and with the variation of $\lambda_E$. Here, the asymptotical dynamics of the $T_{OT}$-component are taken into account. The black line and the dash line represent the stable and the unstable fixed points, respectively. For each $\lambda_E$, the blue and the red dots represent the eventually upper-and-lower boundaries of the stable and the unstable limit cycles in the $T_{OT}$-component. (B) The trajectories of the system (8) when $\lambda = 90.9\,\text{Hz}$ and $n = 22$ (see also the phase orbit in Fig. 5F). The sharp peaks in the left plot and the sharp valleys in the right plot reflect the characteristics of the slow-fast dynamical system. (C) The bifurcation transition regulated by the input rate $\lambda_E$ with $n = 22$. The inner plot indicates the dynamics of the input rate with respect to time. We set $\lambda_E(t) = 57\,\text{Hz}$ for $t \in [0,500]$ (in Second), $\lambda_E(t) = 62\,\text{Hz}$ for $t \in [500,1100]$, $\lambda_E(t) = 200\,\text{Hz}$ for $t \in [1100,1600]$ and $\lambda_E(t) = 90\,\text{Hz}$ for $t \in [1600,2100]$. (D) Network bursting dynamics in: (blue line) the SNN composed of 48 neurons and 12 dendrites. (red line) the 'network' replicated from the traces of voltage and store level in the mean field model with $n = 22$. Note that bursting event is recorded if the firing rate is larger than 30Hz in the SNN, while the stable limit cycle induces the network bursting in the mean field model.

# Table 1. The Model Parameters Used For Simulations

| Name | Description | Value | Units |
|---|---|---|---|
| $N$ | Number of cells | 48 | |
| $n_b$ | Number of bundles | 12 | |
| $\tau$ | Membrane time constant | 10.8 | ms |
| $v_{rest}$ | Resting potential | -62 | mV |
| $a_E(v_E - v_{rest})$ | EPSP amplitude | 4 | mV |
| $a_I(v_{rest} - v_I)$ | IPSP amplitude | 4 | mV |
| $v_E$ | EPSP reversal potential | 0 | mV |
| $v_I$ | IPSP reversal potential | -80 | mV |
| $\lambda_I$ | Inhibitory input rate | 80 | Hz |
| $\tau_{OT}$ | Time decay of oxytocin-induced depolarization | 1 | s |
| $k_{OT}$ | Depolarization for unitary oxytocin release | 0.5 | mV |
| $\Delta$ | Time delay for oxytocin release | 5 | ms |
| $k_p$ | Priming rate | 0.5 | $s^{-1}$ |
| $\tau_r$ | Time constant for priming | 400 | s |
| $k_r$ | Fraction of dendritic stores released per spike (max) | 0.045 | |
| $\tau_{rel}$ | Maximum inter-spike interval for release | 50 | ms |

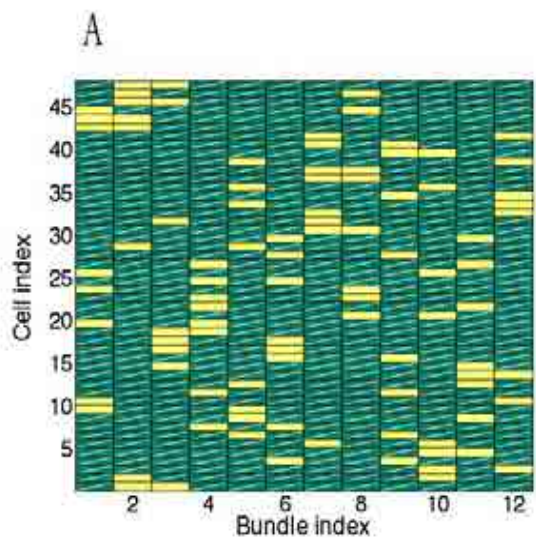
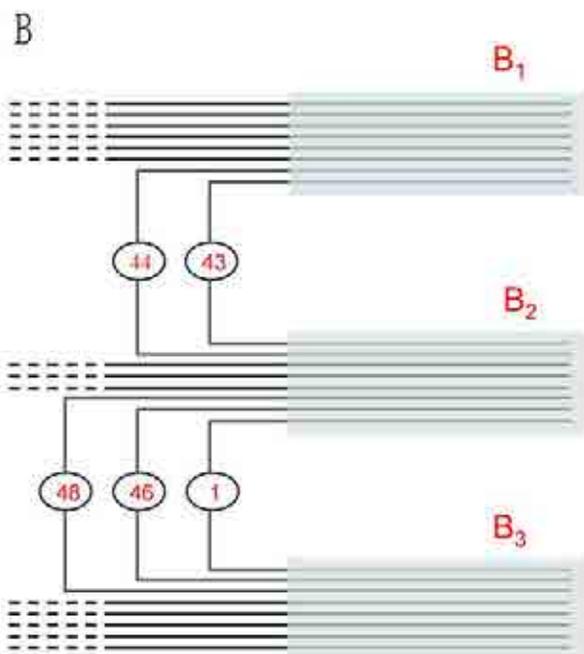
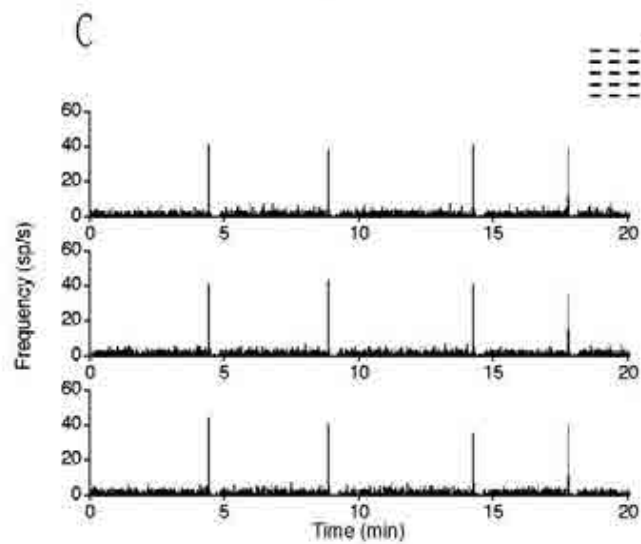
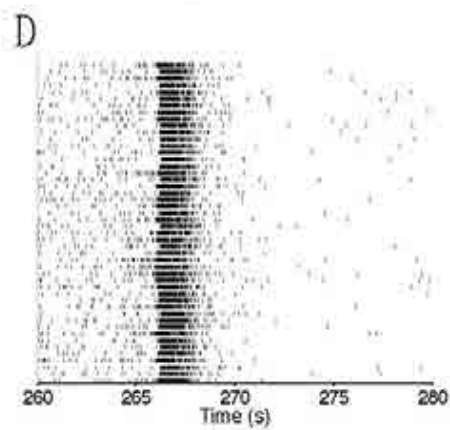
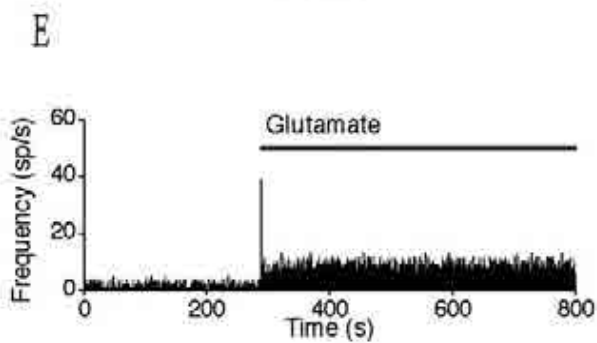
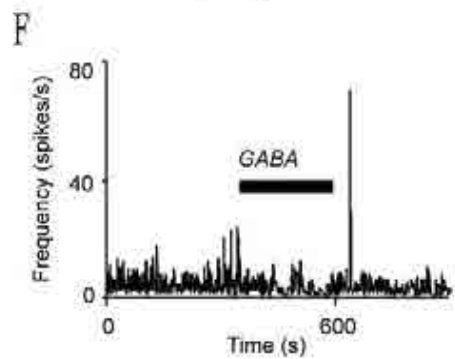

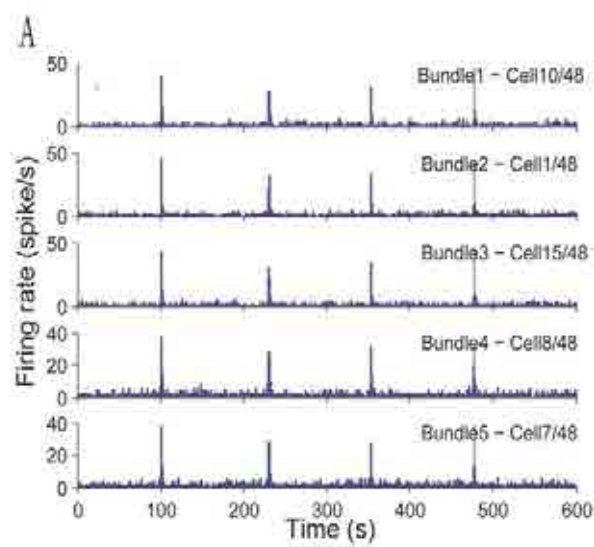 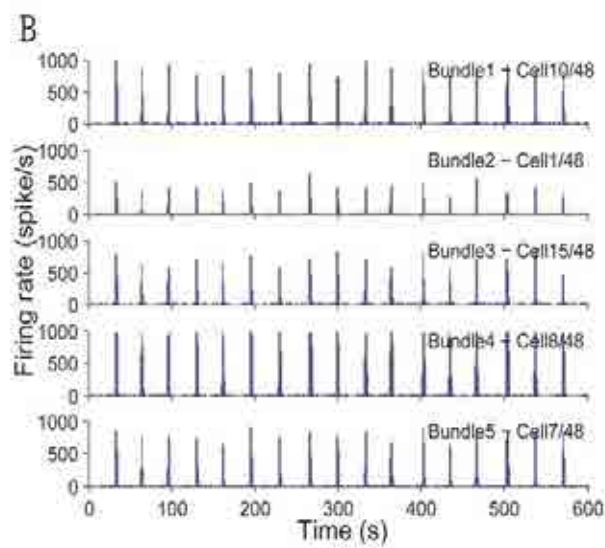
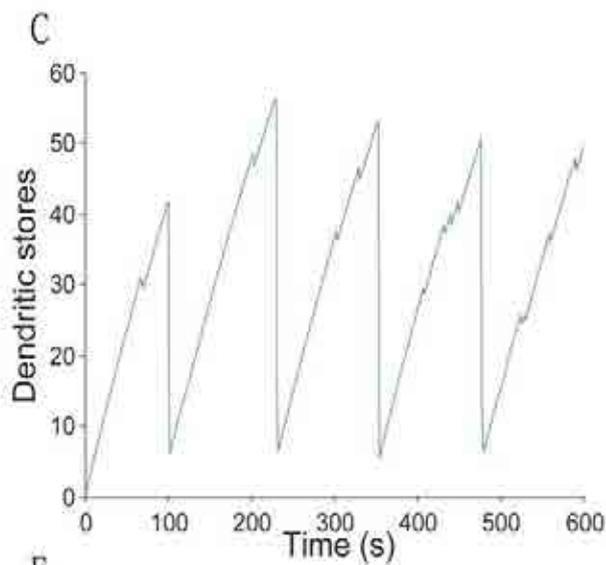 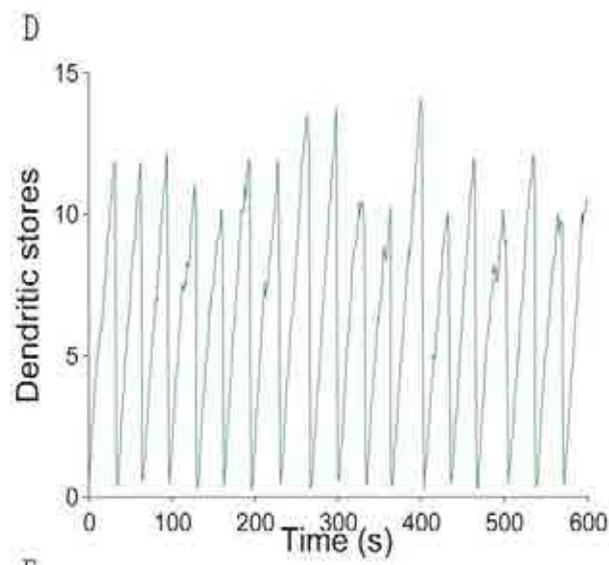
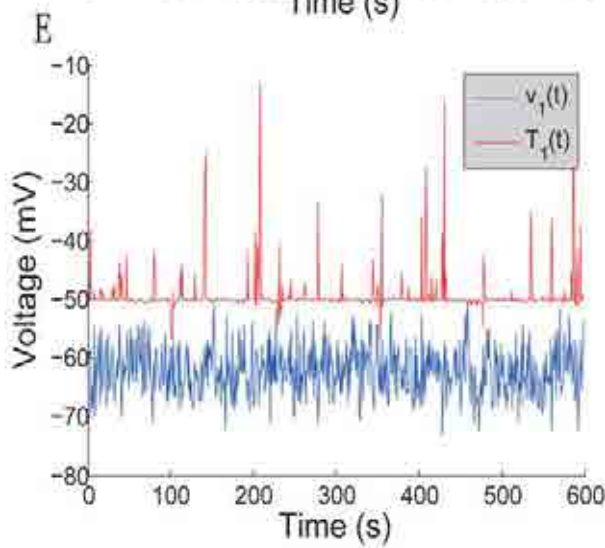 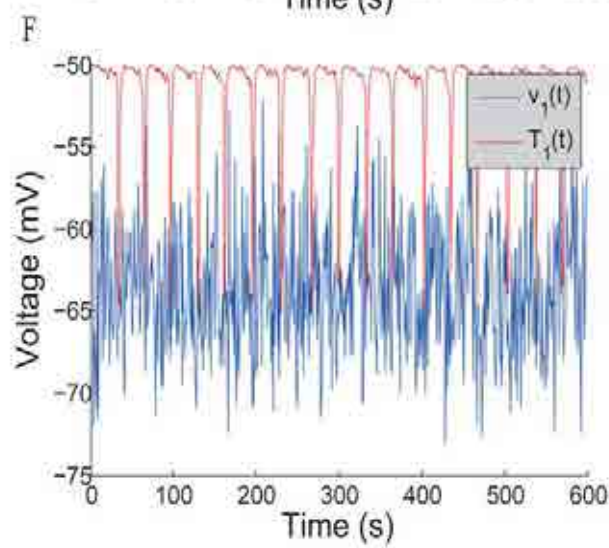

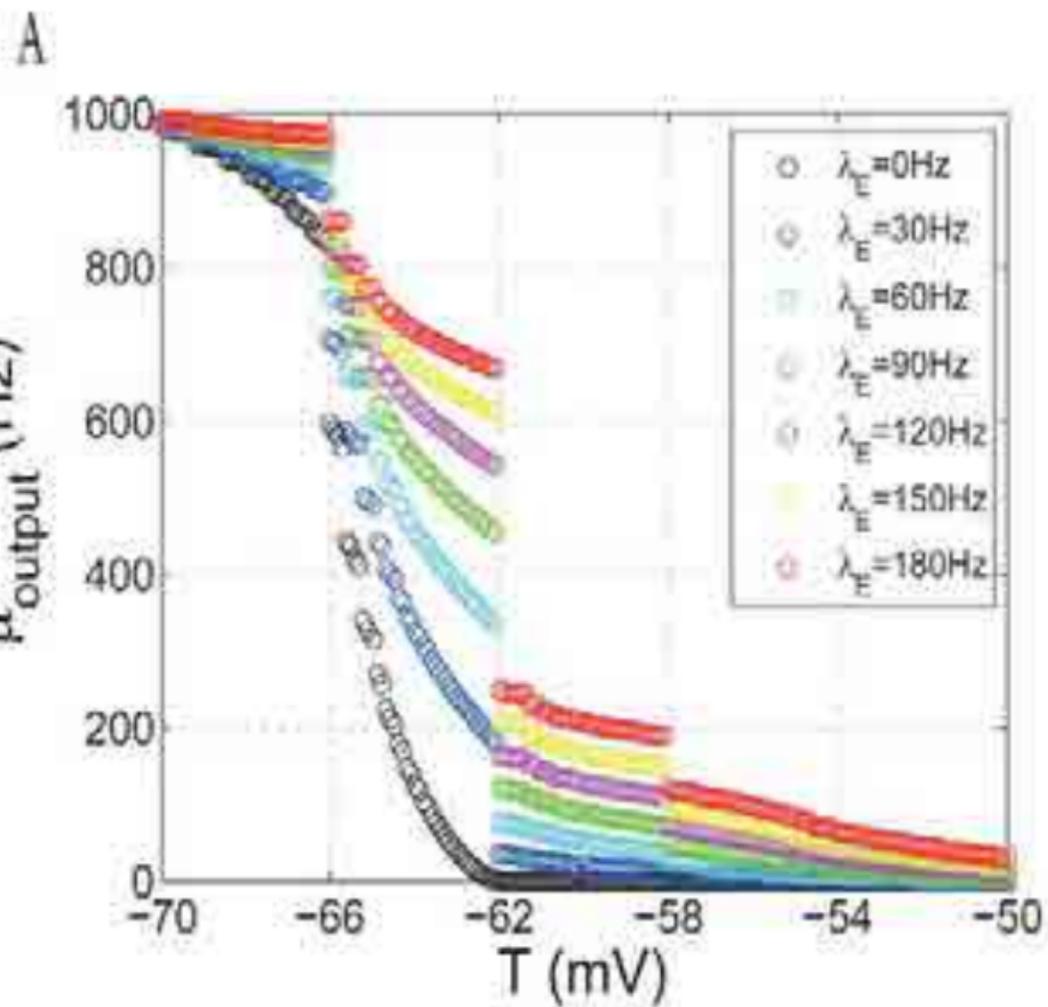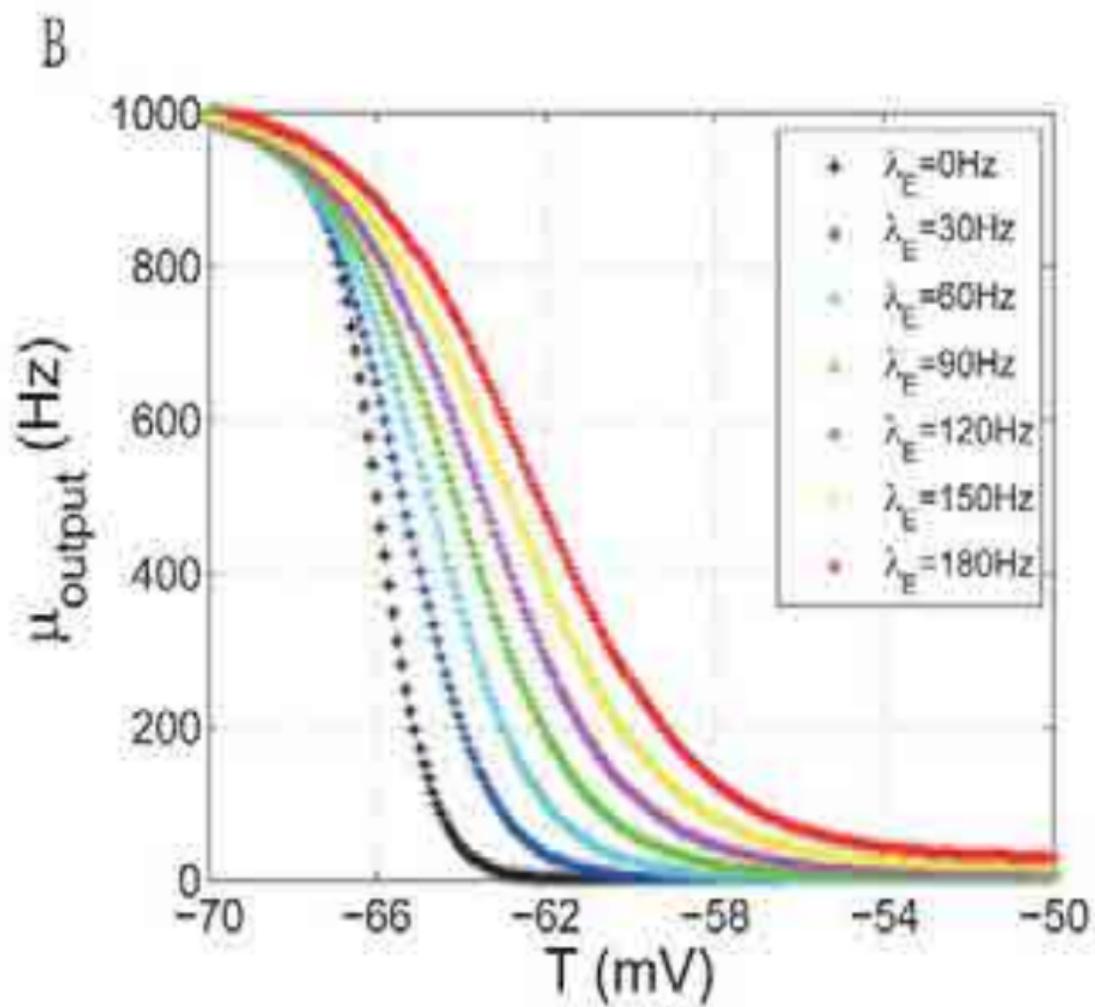

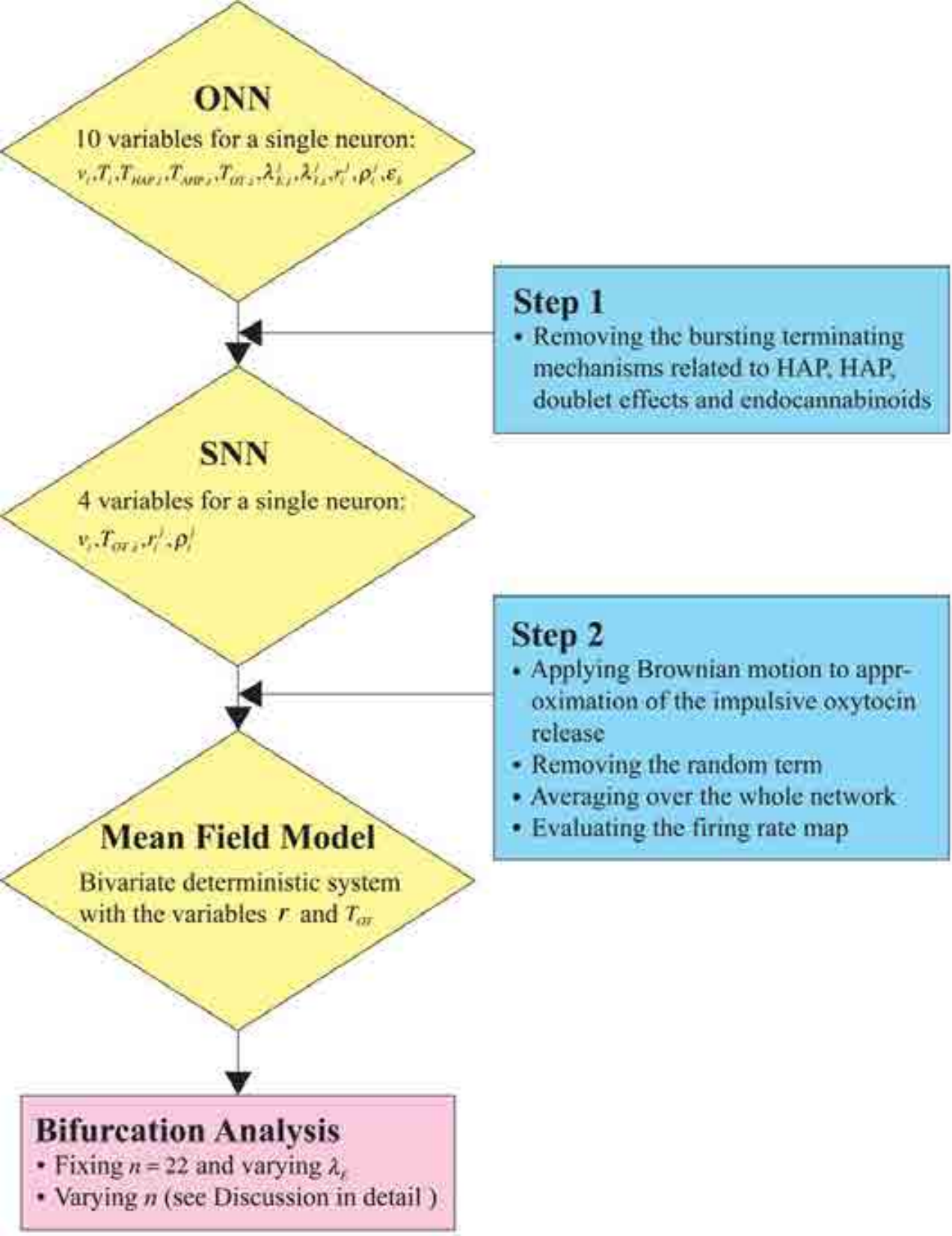

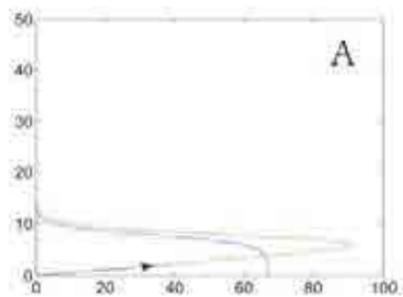
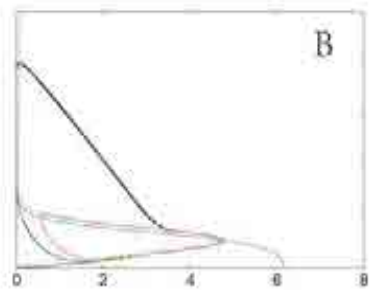
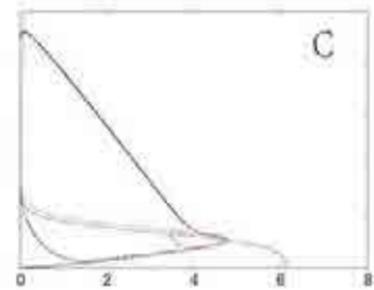
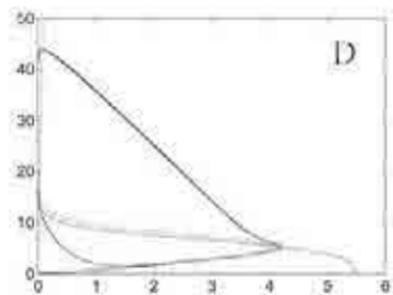
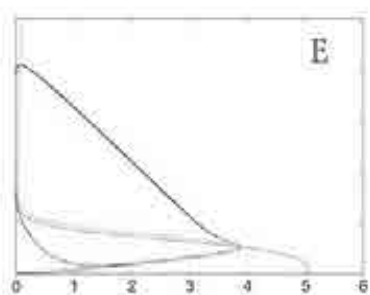
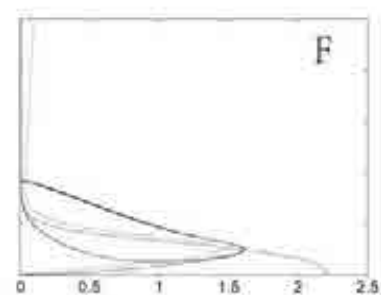
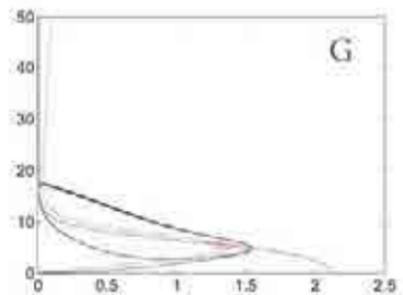
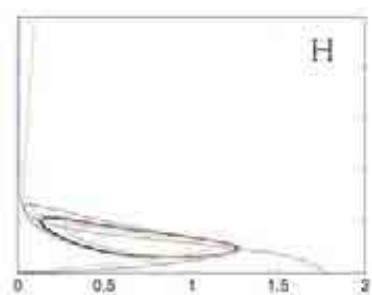
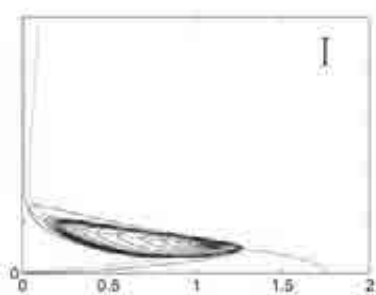

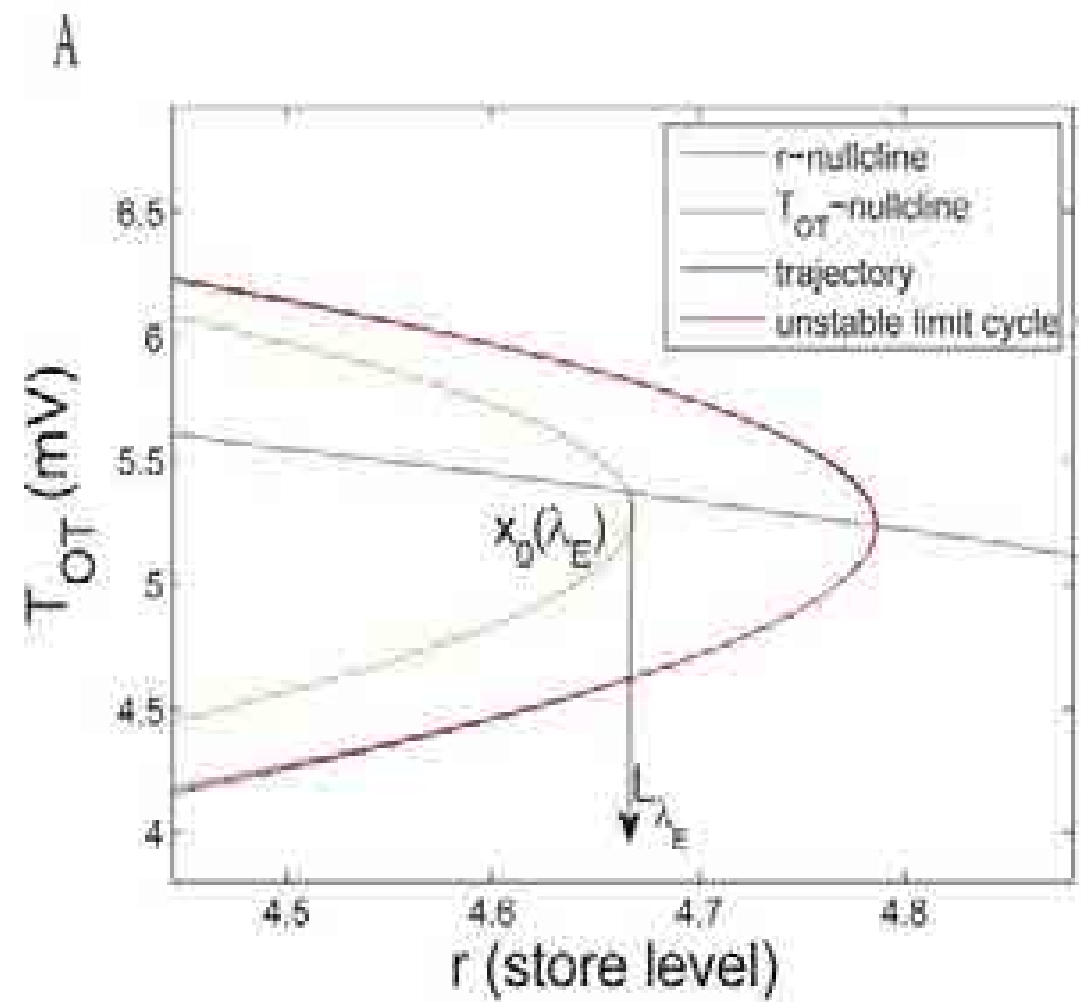 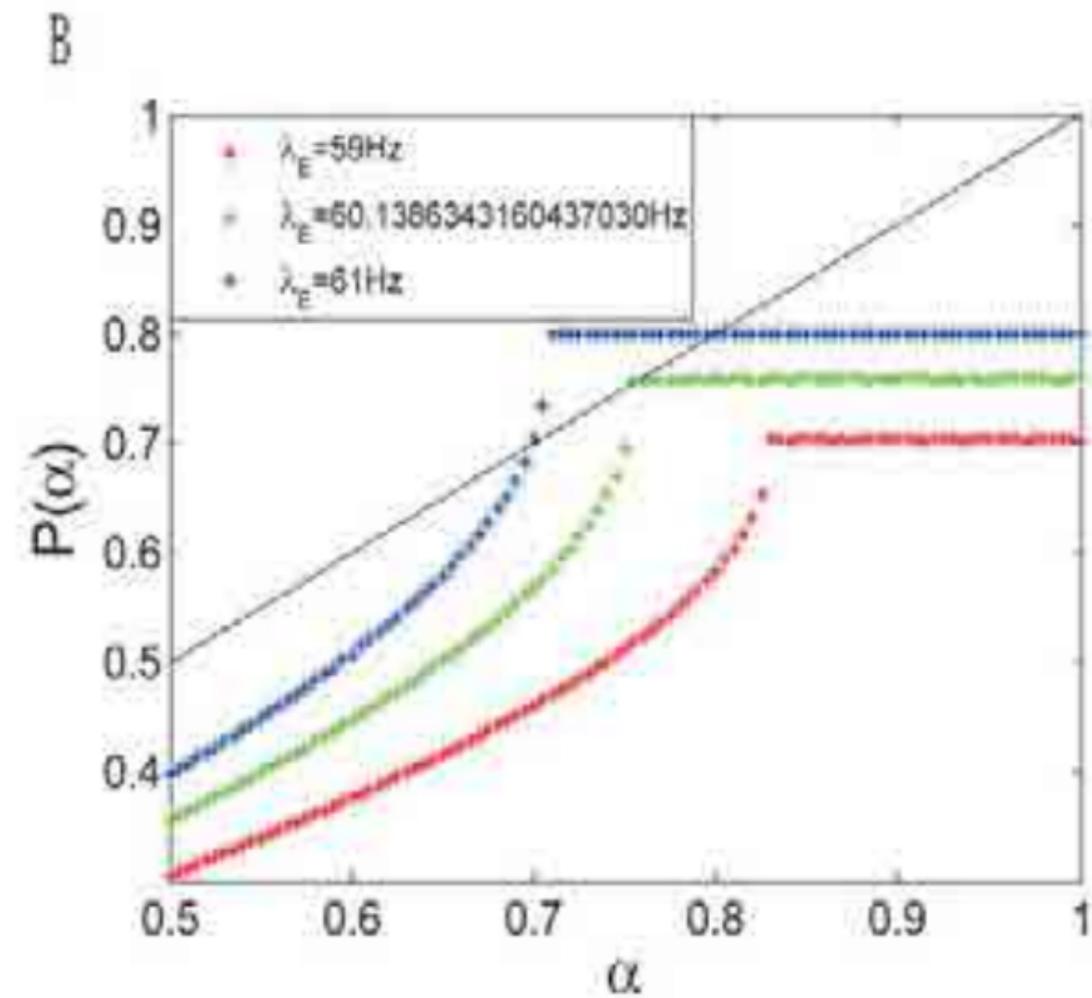

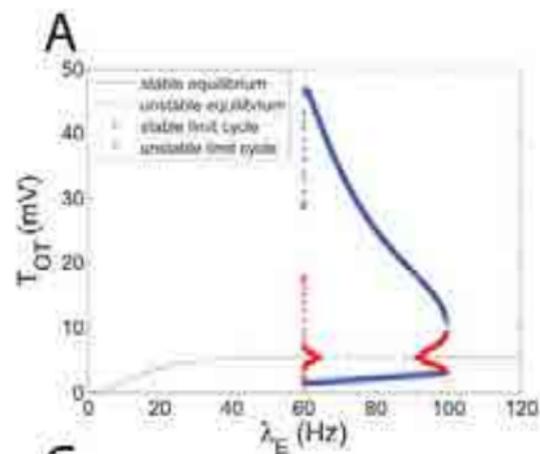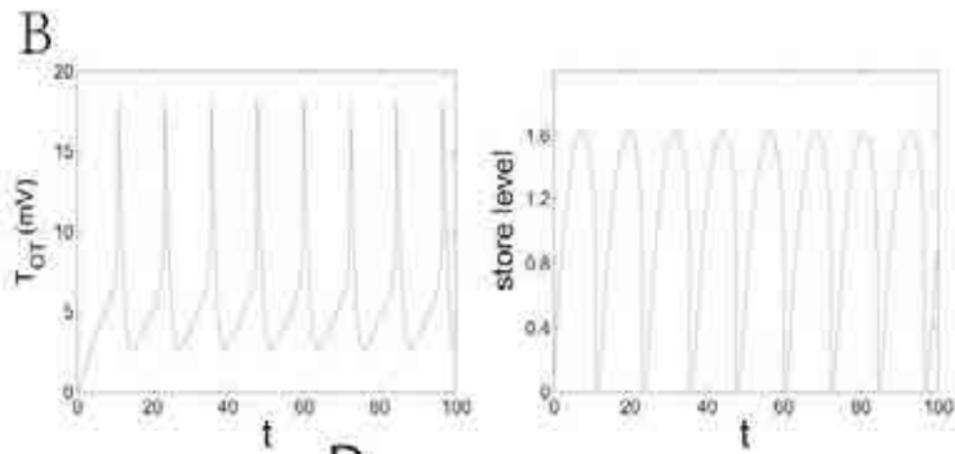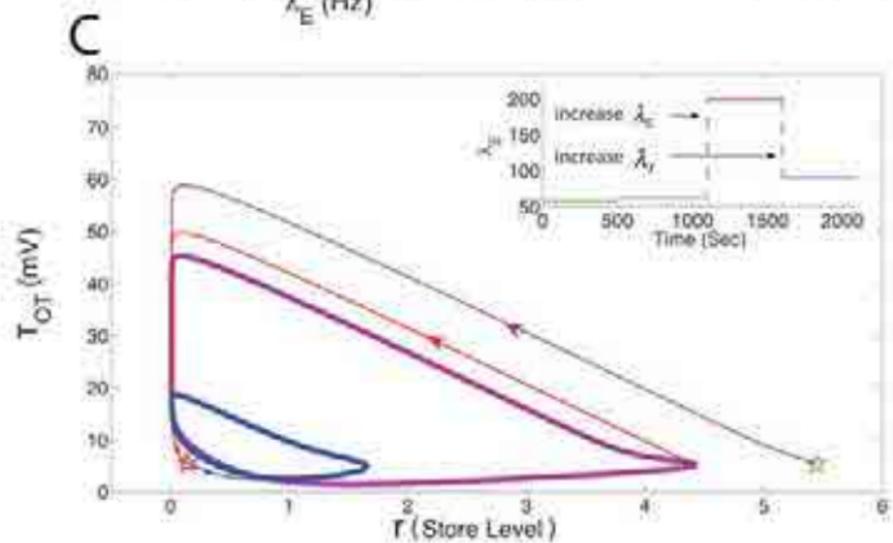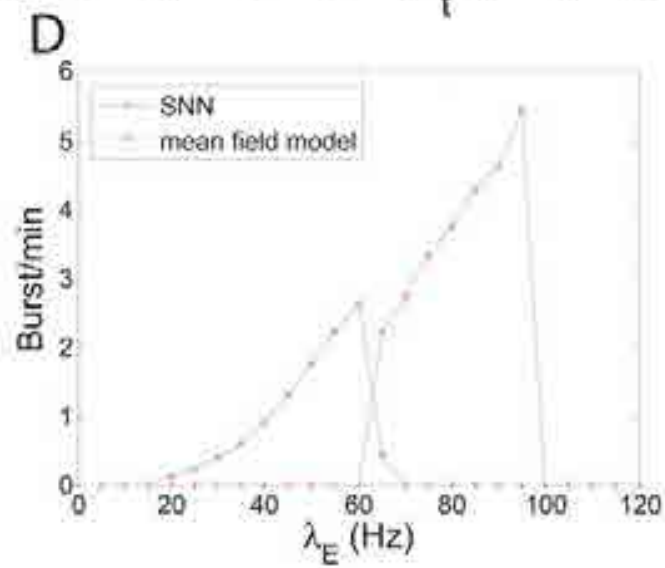